\documentclass[finalversion,simpleeqnnos]{yrl}
\usepackage{times}
\usepackage{graphicx,psfig,epsfig}
\ifx\pdfannotlink\dummy   
\usepackage[hypertex]{hyperref} 
\else
\usepackage[pdftex]{hyperref}   
\fi
\title{{\LARGE\bf 
Constructing a Maximum Utility Slate of On-Line Advertisements}}

\author{
S. Sathiya Keerthi and John A. Tomlin\\
Yahoo! Research\\
2821 Mission College Blvd.\\
Santa Clara, CA 95054\\
{\tt\{selvarak, tomlin\}@yahoo-inc.com}}

\date{April 4, 2007}      

\yrlnumber{YR-2007-001}

\begin{document}             

\maketitle                   

\begin{abstract}
We present an algorithm for constructing an optimal slate of sponsored search advertisements which respects the ordering that is the outcome of a generalized second price auction, but which must also accommodate complicating factors such as overall budget constraints. The algorithm is easily fast enough to use on the fly for typical problem sizes, or as a subroutine in an overall optimization.
\keywords{advertising, budgets, column generation, sponsored search}
\end{abstract}

\section{Introduction}

In this paper we consider the problem of constructing a maximum utility ``slate'' 
of ads for display in response to either a search term, or content page 
requesting ads from a server. We shall treat both of these essentially the 
same---that is that some ordered set of ads (the ``slate'') is to be returned 
in response to a ``query''.

Let us denote the $n$ bidders on this query by
$j=1,\ldots,n$, and suppose that some form of Generalized Second Price auction (GSP)\cite{EOS} has been carried out, which induces a ranking of the bidders. For simplicity, let the numerically ordered indices
also indicate the bid ranking, initially assumed to be determined solely by the bids---sometimes called the Overture ranking\footnote {This assumption will later be generalized.}. That is, if we denote the bid of bidder $j$ by
$A_j$, then 
\[ A_1 \ge  A_2 \ldots \ge A_n. \]
Let $m$ be the maximum number of positions, and let 
$T_{jp}$ be the click-through rate (CTR) of bidder $j$ when his ad is at 
position $p$. Finally, let  $\rho_j$ be a ``utility factor'' associated 
with the appearance of bidder $j$'s ad in response to the query.

We define a slate as an ordered subset $S=\{j_1,\ldots,j_k\}$, where $k\le m$,
of the ordered set of ads $\{1,\ldots,n\}$.
Since we are initially assuming a second-price auction by bid value, bidder $j_p$
in position $p$ pays the bid $A_{j_{p+1}}$ of the bidder occupying position $p+1$.
In addition there is a minimum bid $\epsilon$, which is paid by the last bidder in the slate if and only if there are no lower bidders on the query.
Under these assumptions we wish to solve
\begin{equation}
\max_S\ \  U = \sum_{p=1}^k \rho_{j_p} T_{j_p p} A_{j_{p+1}} 
\label{equ:opt}
\end{equation}
subject to the requirement that $j_1,\ldots,j_k$ is
an increasing set of indices, and $k\le m$.

Assuming the CTRs are independent, the utility $U$ of the slate when the 
$\rho_j$ are all unity is easily seen to be the expected revenue from the slate.
However there may be several reasons why we wish to consider other values of
the $\rho_j$. Some of the more important of these are:
\begin{enumerate}
\item The advertiser placing the ad may be at, or near, its budget, thus reducing the desirability of showing it. This is our primary motivation, and this is a companion paper to \cite{AMT}, which discusses this in detail. For convenience, we include a brief description of this model in an Appendix.
\item ``Ad Fatigue'' induced by too-frequent showing of an ad may reduce its effectiveness. We might therefore wish to penalize some ads which have been shown above some threshold.
\item We may be given (or wish to have) the CTRs as a product two components - a component solely due to the ad, independent of position, and a position-only dependent component. In this case we may reduce $T_{jp}$ to a position only component $T_p$ and an ad-dependent component which becomes a contributor to $\rho_j$. However, this is not a necessary feature and we will usually continue refer to the CTRs as $T_{jp}$.
\end{enumerate}
The treatment we give here is independent of the source of the $\rho_j$.

There is some superficial similarity with the well-known knapsack problem\cite{MT}---we are
selecting a subset $S$ of up to $m$ items (the ads), subject to constraints. Also related is the ``knapsack auction'' considered in \cite{AH}. Even more strongly related is the cardinality constrained knapsack problem\cite{deFN}, since our slates have a fixed maximum size. However, the ordering requirement makes the problem more specialized. Nevertheless, like the knapsack problem, our problem is amenable to a dynamic programming approach.

\section{Backward Recursion}

We begin by giving a $O(n^2m)$ algorithm for solving (\ref{equ:opt})
using a dynamic programming (DP) algorithm with backward recursion\cite{Bell}.
For this approach, it is convenient to include $m$ dummy
bidders, call them $n+1,\ldots,n+m$, with $\rho_j=0$ and $A_j=\epsilon$
for $j=n+1,\ldots,n+m$. The reason for doing this is that we only need
to consider slates of size equal to $m$; smaller slates can be 
padded with dummy bidders at the end to produce the same effect and
make a slate of size exactly $m$.

Take one $j$ and one $s$ such that $1\le s\le m$. Let us define a
{\em subslate} starting from $j$ at position $s$ as a set of increasing
indices, $\tilde{S}=\{j_s,j_{s+1},\ldots,j_m\}$ such that $j_s=j$.
Let $\tilde{\cal{S}} (s,j)$ denote the set of all such subslates. Consider the
problem of computing the best ``marginal-revenue-to-go":
\begin{equation}
F(j,s)=
\max_{\tilde{S}\in\tilde{\cal{S}} (s,j)} \sum_{p=s}^m \rho_{j_p} T_{j_p p} A_{j_{p+1}} 
\label{equ:dpopt}
\end{equation}
Suppose we fix $j_{s+1}$ and proceed optimally from there. If $F(j_{s+1},s+1)$
is known for all $j_{s+1}>j$
then we can compute (\ref{equ:dpopt}) in standard dynamic programming fashion as:
\begin{equation}
F(j,s)=
\max_{j_{s+1}>j} \rho_j T_{js} A_{j_{s+1}} + F(j_{s+1},s+1)
\label{equ:dpform}
\end{equation}
We can start the DP algorithm by setting $F(j,s)=0$ $\forall$ 
$j=n+1,\ldots,n+m$, $s=1,\ldots,m$. Now recurse backwards and compute
$F(j,s)$ $\forall$ $s=1,\ldots,m$ and $j=n,\ldots,1$. Finally, choose
$\max_j F(j,1)$ to get the solution of (\ref{equ:opt}) as well as
the optimal slate.

\section{An Optimal Path Approach}

%
%
\begin{figure}[htfb]
\begin{center}
\hspace{0.0cm}
\psfig{figure=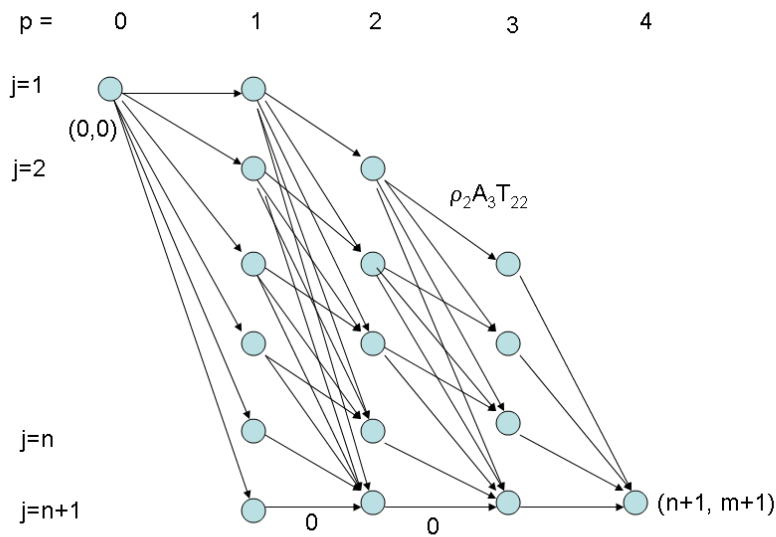,width=5in}
\caption{Network with $n>m$}
\label{fig:network1}
\end{center}
\end{figure}
%
%
\begin{figure}[htfb]
\begin{center}
\hspace{0.0cm}
\psfig{figure=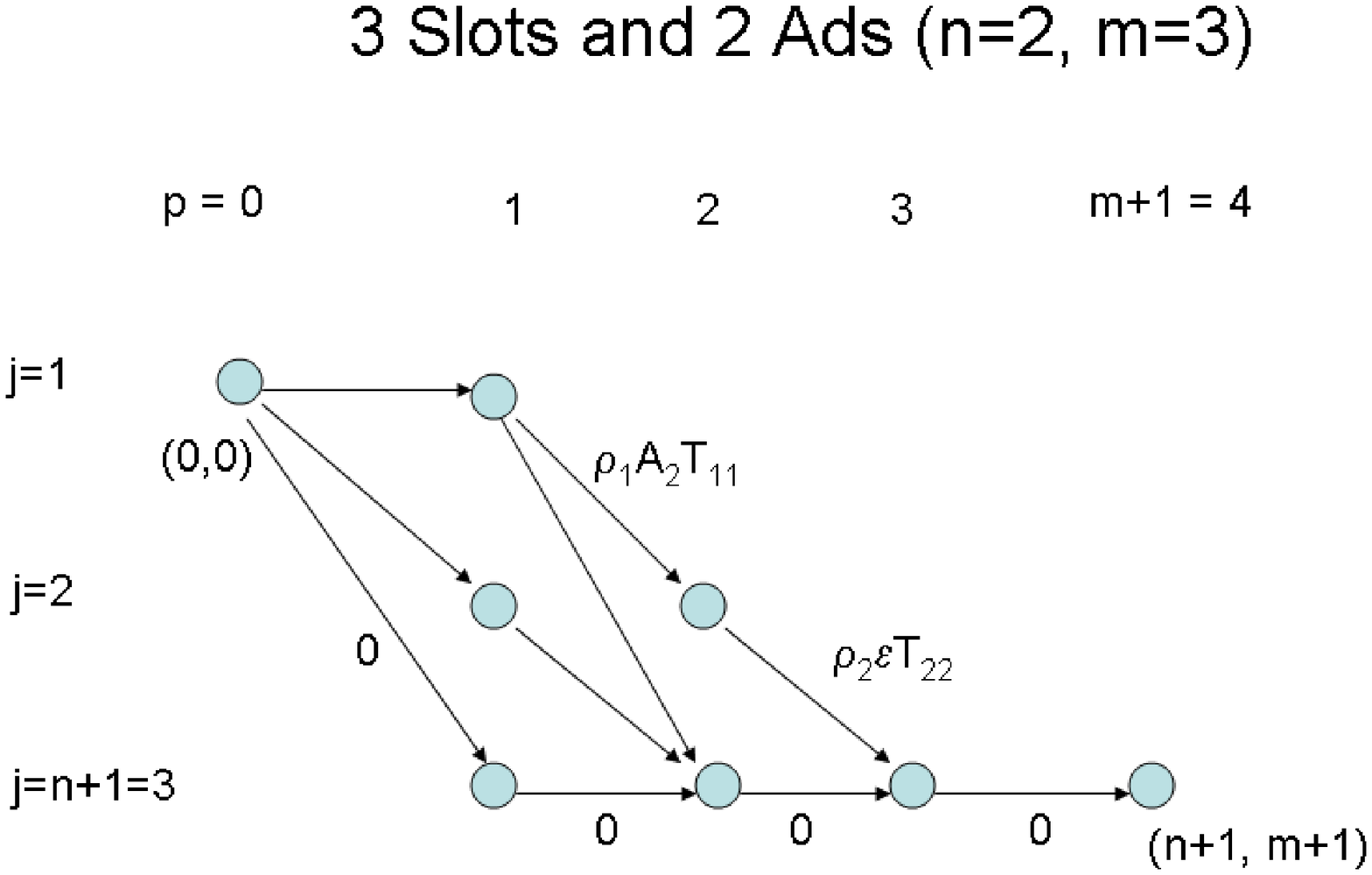,width=5in}
\caption{Network with $n<m$}
\label{fig:network2}
\end{center}
\end{figure}

Frequently, problems that are amenable to dynamic programming can be cast in the form of a shortest or longest path problem \cite{Be0}, and this is no exception. Let us define a network with nodes $N_{j,p}$ for $p=1,...,m$ and $j = p,...,n+1$. We also define terminal nodes $N_{0,0}$ and $N_{n+1,m+1}$. The directed edges and their associated costs are defined as:
\begin{eqnarray*}
\begin{array}{llll}
(N_{0,0}, N_{j,1})&:&c_{0,j,0} = 0\\
& &(j = 1,...,n+1) \\
(N_{i,p}, N_{j,p+1})&:& c_{i,j,p} = \rho_iA_jT_{ip}\\
& &(j>i\ge p= 1,...,m-1) \\
(N_{j,m}, N_{n+1,m+1})&:& c_{j,n+1,m} = \rho_jA_{j+1}T_{jp}\\
& &(j = p,...,n-1)  \\
(N_{n,m}, N_{n+1,m+1})&:& c_{n,n+1,m} = \rho_n\epsilon T_{nm}  \\
(N_{n+1,p}, N_{n+1,p+1})&:& c_{n+1,n+1,p} = 0\\
& &(p=1,...,m)
\end{array}
\end{eqnarray*}
where $c_{i,j,p}$ is the cost for the edge directed from $N_{i,p}$ to $N_{j,p+1}$ and $\epsilon$ is the minimum bid as before. Note that not all these edges are defined (or need to be defined) if $n<m$. (See figures 1 and 2).

Since the network is directed and acyclic, and the utilities/distances associated with the non-trivial edges correspond to the utility of placing ad $i$ in slot $p$, {\em followed by} ad $j$ in slot $p+1$,
it is easy to see that the longest path from $N_{0,0}$ to $N_{n+1,m+1}$ maximizes (\ref{equ:opt}) - or equivalently the shortest path using the negatives of the costs defined above. 

Very efficient algorithms are known for the shortest path problem, but in our case the problem is small, so a simpler implementation suffices. 

Since the forward recursion/ optimal path approach is more intuitive and visually appealing, we shall use it for the remainder of this paper in the discussion of extensions and variations.

\section{Extension to Revenue Ranking}

The present scheme extends to what is sometimes known as revenue ranking, where the ads are ranked not just by bid, but by ``expected revenue'' which is modeled as the product of the advertiser's bid $A_j$ and $Q_j$, which is the ``quality score'', or ``clickability'', for bidder $j$'s ad for this query. This quantity is thought to better represent the value of a bidded ad than the raw bid.

The ads are now ranked according to this product, so that:
\[ A_1Q_1 \ge  A_2Q_2 \ldots \ge A_nQ_n. \]
To preserve the condition that the expected payment for a click is at least that of the next ranked bidder, we require that the expected cost per click (CPC) of bidder $j_p$ must be at least $A_{j_{p+1}}\frac{Q_{j_{p+1}}}{Q_{j_p}}$.
Using this observation, the path technique extends to the expected revenue ranking scheme. In this case the objective function is now:
\begin{equation}\label{obj2}
{\rm Maximize}\ \ \tilde{U} = \sum_{p=1}^m \rho_{j_p}A_{j_{p+1}}\frac{Q_{j_{p+1}}}{Q_{j_p}}T_{j_p p}
\end{equation}
where  The network model above remains the same except that the edge costs are now modified to be:
\begin{eqnarray*}
\begin{array}{llll}
(N_{0,0}, N_{j,1})&:&c_{0,j,0} = 0\\
& &(j = 1,...,n+1) \\
(N_{i,p}, N_{j,p+1})&:& c_{i,j,p} = \rho_iA_j\frac{Q_j}{Q_i}T_{ip}\\
& &(j>i\ge p= 1,...,m-1) \\
(N_{j,m}, N_{n+1,m+1})&:& c_{j,n+1,m} = \rho_jA_{j+1}\frac{Q_{j+1}}{Q_j}T_{jp}\\
& &(j = p,...,n-1)\\
(N_{n,m}, N_{n+1,m+1})&:& c_{n,n+1,m} = \rho_n\epsilon T_{nm}  \\
(N_{n+1,p}, N_{n+1,p+1})&:& c_{n+1,n+1,p} = 0\\
& &(p=1,...,m) 
\end{array}
\end{eqnarray*}

\section{Further Extensions}

The path technique extends to other practically useful variants. Two of these include introducing restrictions on the subset of ads which may be omitted from the slate, and use of a hybrid objective function which is made up of a weighted sum of the first and second price utilities.

%
%
\begin{figure}[htfb]
\begin{center}
\hspace{0.0cm}
\psfig{figure=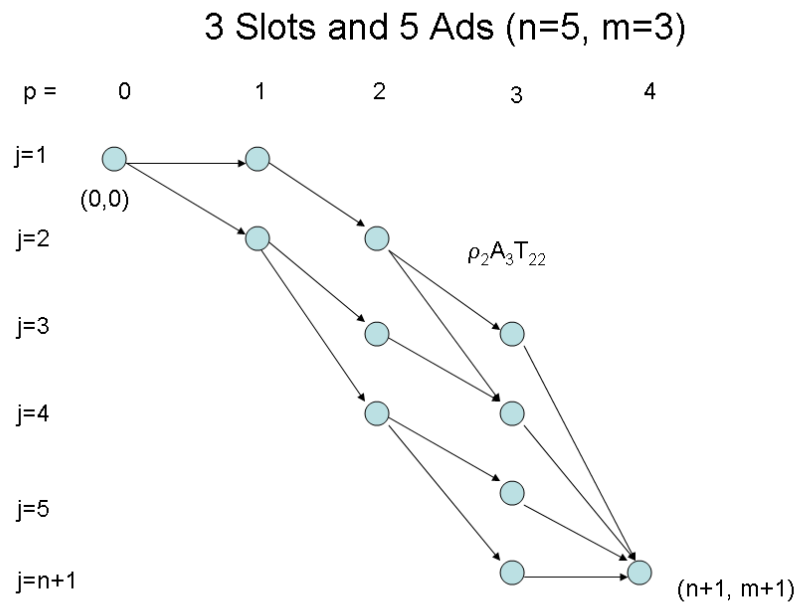,width=5in}
\caption{Reduced network with $mask=(1,0,1,0,1)$}
\label{fig:network3}
\end{center}
\end{figure}

\subsection{Restricted Omissions}
The algorithm(s) we have been considering assume that any appropriate ordered subset of the ads may be chosen which fits within the slate size. In practice this may not always be true. While it is obviously legitimate to exclude ads from the limited space when the bids  (or expected revenues) are too low, it is not so obvious that this may be done for other reasons. For example, one of the motivations we cited for using non-unit weights $\rho_j$ was the need to accommodate bidders with limited budgets. One means of doing this is to allow budgeted bidders to be held out of the notional auction---that is excluded from the slate. However, it is not obvious that this option should extend to unbudgeted bidders. This must be a business decision. We therefore require a means of specifying which ads (bidders) can be excluded for reasons other than low rank. We accomplish this by specifying a {\em mask} or bit vector, which has a 1 if the ad can be excluded and a zero otherwise. We then modify the algorithm as follows:

Since each arc in the network gives the utility of including a particular ad $i$ in position $p$ followed by ad $j$, we consider only arcs such that:
\begin{enumerate}
\item For each position $p$ we allow $i$ to assume the values from $p$ up to the first ad in rank order which has a zero mask bit. Any subsequent ads are ignored for this $p$. This ensures that the unmasked ad with the highest rank is not excluded, but that lower ranked ads which are masked are not considered for the
position $p$.
\item For each $i$ chosen as above, the second index $j$ shall only run from
$i+1$ through the next unmasked ad. This ensures that if an ad $i$ can be followed by an unmasked ad it will be the next in rank order.
\end{enumerate}
This scheme may be thought of as actually removing arcs from the networks such as those shown in the figures, or more simply implemented by modifying the longest path algorithm with the rules we have itemized. In Figure \ref{fig:network3} we show the reduced network obtained from Figure \ref{fig:network1} when we specify that $mask = (1,0,1,0,1)$. In practice however, we use the second technique of modifying the algorithm.

\subsection{Hybrid Objectives}

Thus far we have assumed some form of generalized second price auctions is implicit in the utility of the slate. However, even if the price per click is computed with such an assumption, we may wish to include other factors in our utility calculation. For example we may wish to consider the first prices, on the assumption that in a truthful setting these are the actual values placed by the bidders on a click for their ad and we wish to take this into account. Alternatively we may be interested in the raw number of expected clicks. Both of these situations can be accommodated by considering a composite weighted objective function which takes into account both first and second prices in specifying the arc costs in the longest path algorithm. Let us define the hybrid objective as:
\begin{equation}\label{objx}
{\rm Maximize}\ \ \hat{U} = \sum_{p=1}^m \mu_{j_p}A_{j_p}T_{j_p p} + \sum_{p=1}^m \rho_{j_p}A_{j_{p+1}}\frac{Q_{j_{p+1}}}{Q_{j_p}}T_{j_p p}
\end{equation}
which may be re-written:
\begin{equation}\label{objy}
{\rm Maximize}\ \ \hat{U} = \sum_{p=1}^m (\mu_{j_p}A_{j_p} + \rho_{j_p}A_{j_{p+1}}\frac{Q_{j_{p+1}}}{Q_{j_p}}  )T_{j_p p}
\end{equation}
Then if we rewrite the relevant arc costs as:
\begin{eqnarray*}
\begin{array}{llll}
c_{i,j,p} &=& (\mu_{i}A_{i}+\rho_iA_j\frac{Q_j}{Q_i})T_{ip}\\
& &(j>i\ge p= 1,...,m-1) \\
c_{j,n+1,m} &=& (\mu_{j}A_{j}+\rho_jA_{j+1}\frac{Q_{j+1}}{Q_j})T_{jp}\\
& &(j = p,...,n-1) \\
\end{array}
\end{eqnarray*}
we may solve the optimal path problem as before. 

This framework is very flexible and can lead to many extensions. For example if we wish to have a weighted combination of expected revenue and expected clicks we may set the $\mu_{j}$ to $1/A_{j}$.

Our colleague Zo\"e Abrams has also informed us\cite{Zoe} that the dynamic programming approach to column generation extends to the Vickery-Clark-Groves (VCG) auction mechanism.

\section{Computational Results}

All of the algorithms we have described are efficient in terms of number of operations, especially since the numbers involved are relatively small in the on-line advertising framework. Typically $m$ is less than or equal to about 12, and the number of bidders $n$ which need to be considered for inclusion in a slate is less than 100, and may even be less than $m$.

We have implemented the forward algorithm in a straightforward way in C++, to be called as a subroutine in the column generation algorithm of \cite{AMT}. When run on a 32-bit Linux box with a 2.8 GHz Xeon processor, it takes an average of 25 microseconds for a sample of 5000 queries where there are 12 ad positions, and between 1 and 77 candidate ads. This includes setting up the data structures as well as the actual path computation, and is obtained without any attempt to optimize the code other than using the {\tt -O2} option of the {\tt gcc} compiler.
 We can therefore afford to execute the algorithm repetitively, either in real time in an on-line setting, or repeated many times as a subroutine.

\section{Conclusion}

Sponsored search auctions have recently received considerable attention, but the subsequent problem of how to implement, or adapt, the outcomes in the presence of complicating factors such as budget constraints appears to have been less well studied. We have shown that this can be accomplished in cases of practical interest by a simple but very fast dynamic programming algorithm.

\bibliographystyle{abbrv}
\bibliography{utility} 

\section*{Appendix}

The algorithm(s) described in this paper were developed as a column generating subroutine for the linear programming model (LP) described in \cite{AMT} for optimizing sponsored search ad delivery subject to budget constraints. The concept of a ``slate'' is put forward in that paper corresponding to columns of the LP. We may formally state the LP as follows:
\eject
{\sl Indices}
\vskip 5pt
   \begin{tabular}{ll}
     $i = 1,...,N$ &  The queries
   \\$j = 1,...,M$ &  The bidders
   \\$k = 1,...,K_i$ &  The slates (for query $i$)
   \end{tabular}
\vskip 5pt
{\sl Data}
\vskip 5pt
   \begin{tabular}{ll}
    $ d_j$    & The total budget of bidder $j$
   \\$v_i$ & Expected number of occurrences of 
\\ & keyword $i$
   \\ $a_{ijk}$ & Expected cost to bidder $j$ if slate $k$ 
\\ & is shown for keyword $i$
   \\ $r_{ik}$ & Objective function coefficient for slate $k$
\\ & for keyword $i$
\end{tabular}
\vskip 5pt
{\sl Variables}
\vskip 5pt
   \begin{tabular}{ll}
      $x_{ik}$   &  Number of times to show slate $k$ for keyword $i$
   \end{tabular}
\vskip 5pt

{\sl Constraints}
\vskip 5pt
(Budget)
 \begin{equation} \label{eqn-bud}
	\sum_i\sum_k a_{ijk} x_{ik} \le d_j \ \ \ \ \ \ \forall j
 \end{equation}
\vskip 5pt
(Inventory)
 \begin{equation} \label{eqn-Inv}
       \sum_k x_{ik} \le v_i \ \ \ \ \ \ \forall i
 \end{equation}
\vskip 5pt

{\sl Objective}
\vskip 5pt
   \begin{tabular}{ll}
    {\rm Maximize} &$ \ \sum_{i}\sum_k r_{ik} x_{ik}$
   \end{tabular}
\vskip 5pt

Each column of the LP, that is the $a_{ijk}$ values, corresponds to the expected cost per click (CPC) to budgeted advertisers if their ad is clicked on when slate $k$  is shown for query $i$. If there are more then a handful of budgeted bidders for a query, the number of possible slates is enormous. We therefore require a method for generating those columns which may be included in the LP optimum solution (the well-known idea of ``column generation''\cite{LD}). When the objective function coefficients are the expected revenue from a slate (i.e. $r_{ik} = \sum_j a_{ijk}$), and the dual values corresponding to the budget constraints are $\pi_j$, the subproblem we wish to solve is of the form (\ref{obj2}) with $\rho_j = 1 - \pi_j$. If the objective coefficients are the bid values (assumed to reflect a bidders true value for a click), the subproblem is of the form (\ref{objx}) with $\mu_j = 1$ and $\rho_j = -\pi_j$. In either case, the slate generated can potentially improve the LP solution if the value is greater than the dual value (say $\gamma_i$) for the $i^{th}$ query volume constraint . See \cite{AMT} for much greater detail.

\end{document}